\documentclass[sn-nature]{sn-jnl}

\usepackage{graphicx}%
\usepackage{multirow}%
\usepackage{amsmath,amssymb,amsfonts}%
\usepackage{amsthm}%
\usepackage{mathrsfs}%
\usepackage{physics}
\usepackage[title]{appendix}%
\usepackage{xcolor}%
\usepackage{textcomp}%
\usepackage{manyfoot}%
\usepackage{booktabs}%
\usepackage{algorithm}%
\usepackage{algorithmicx}%
\usepackage{algpseudocode}%
\usepackage{listings}%
\usepackage{lineno}

\theoremstyle{thmstyleone}%
\theoremstyle{thmstyletwo}%

\theoremstyle{thmstylethree}%

\begin{document}

\title[Probing the quantum metric of 3D topological insulators]{Probing the quantum metric of 3D topological insulators}

\author*[1]{\fnm{Giacomo} \sur{Sala}} \email{giacomo.sala@unige.ch} \equalcont{These authors contributed equally to this work.}

\author*[2]{\fnm{Emanuele} \sur{Longo}} \email{elongo@icmab.es} \equalcont{These authors contributed equally to this work.}

\author[3]{\fnm{Maria Teresa} \sur{Mercaldo}}

\author[1]{\fnm{Stefano} \sur{Gariglio}}

\author[4]{\fnm{Mario} \sur{Cuoco}}

\author[5]{\fnm{Roberto} \sur{Mantovan}}

\author*[3]{\fnm{Carmine} \sur{Ortix}} \email{cortix@unisa.it}

\author*[1]{\fnm{Andrea D.} \sur{Caviglia}} \email{andrea.caviglia@unige.ch}

\affil*[1]{\orgdiv{Department of Quantum Matter Physics}, \orgname{University of Geneva}, \city{Geneva}, \country{Switzerland}}

\affil[2]{\orgname{Institute of Materials Science of Barcelona (ICMAB-CSIC)}, \city{Bellaterra}, \country{Spain}}

\affil[3]{\orgdiv{Dipartimento di Fisica ‘E. R. Caianiello'}, \orgname{Universitá di Salerno}, \city{Fisciano}, \country{Italy}}

\affil[4]{\orgdiv{CNR-SPIN c/o Universita’ di Salerno}, \city{Fisciano}, \country{Italy}}

\affil[5]{\orgdiv{CNR-IMM}, \city{Agrate Brianza}, \country{Italy}}


\abstract{The surface states of 3D topological insulators possess geometric structures that imprint distinctive signatures on electronic transport. A prime example is the Berry curvature, which controls, \textcolor{black}{for instance}, electric frequency doubling via \textcolor{black}{its} higher order moments. In addition to the Berry curvature, topological surface states are expected to exhibit a nontrivial quantum metric, which plays a key role in governing nonlinear magnetotransport. However, its manifestation has yet to be experimentally observed \textcolor{black}{and controlled} in 3D topological insulators. Here, we provide evidence for a nonlinear response activated by the quantum metric of the topological surface states of Sb$_2$Te$_3$. We measure a time-reversal odd, nonlinear magnetoresistance that is independent from the temperature and \textcolor{black}{the scattering time} below 30 K, and is thus of intrinsic geometrical origin. \textcolor{black}{This quantum metric magnetoresistance can be controlled by tuning the contributions of the top and bottom topological surface states by voltage gating.} Our measurements thus demonstrate the existence \textcolor{black}{and tunability} of quantum 
geometry-induced transport in topological phases of matter  and provide strategies for designing novel functionalities in topological devices.}


\maketitle

Three-dimensional strong topological
insulators are materials with 
anomalous gapless surface states mandated by
the nontrivial topological properties of the insulating bulk \cite{Hasan2010,Qi2011}. These conductive states 
evade the fermion doubling theorem \cite{NIELSEN1981219} and form
an odd number of topologically-protected Dirac cones that display remarkable features such as spin-momentum locking, robustness against disorder-mediated electronic localization, and suppressed backscattering. These properties underscore the enormous technological potential of topological insulators \cite{Tian2017,Breunig2021} with applications in topological superconductivity \cite{Sato2017} and quantum computing \cite{He2019a} as well as metrology \cite{Huang2025}, spintronics \cite{Pesin2012,He2019a}, nonlinear electronics \cite{He2021}, and energy harvesting \cite{Xu2017a}. 

Topological surface states display transport effects such as the weak antilocalization that originate from the Berry phase \cite{Bardarson2013}, a geometrical property of electronic wavefunctions.
However, Dirac surface states are also equipped with a nontrivial metric tensor, known as quantum metric, that
has recently come to the fore as defining property of the ground state and response functions of quantum materials \cite{Gao2025,Verma2025}. The quantum metric 
encodes electric field-induced corrections to the electron energies and electronic Berry curvature
and is thus pivotal in nonlinear transport and nonlinear optics \cite{Ahn2020,Ahn2022,Jiang2025}.  
More generally, the quantum metric plays a fundamental role in a variety of condensed-matter phenomena such as flat-band superconductivity, magnetism, fractional Chern insulators, and electron-phonon interactions, to 
name just a few
\cite{Gao2025,Verma2025,Yu2025,Liu2024c}. Yet, experimental observations of quantum metric driven effects remain limited to a handful of materials \cite{Tian2023,Gao2023,Wang2023c,Kim2025,Kang2024,Sala2025,Zhao2025c} and are notably lacking in the important class of 3D nonmagnetic topological insulators, where the combination of surface states and quantum geometry may enable new electronic functionalities \cite{Breunig2021}.

Like the Berry curvature, the quantum metric peaks near band degeneracies, where the Bloch states vary most rapidly in momentum space. More importantly, topology is a sufficient, albeit not necessary, condition for quantum geometry \cite{Verma2025}. Gapless Dirac cones on the surfaces of 3D topological insulators are thus expected to host prominent quantum metric effects \cite{Chen2024c,Mera2022}, but no such evidence has been provided so far. Here, guided by recent theoretical predictions \cite{Mercaldo2024}, we probe via nonlinear magnetotransport the quantum metric of the surface states of topological Sb$_2$Te$_3$ thin films. We measure a second-order, nonlinear response that is independent of disorder at temperatures below 30 K and is thus indicative of a surface intrinsic transport of geometric origin. \textcolor{black}{The magnitude of the nonlinear response can be further controlled by tuning the electronic filling of the Dirac cone in field-effect devices.} Our findings demonstrate the existence \textcolor{black}{and tunability} of quantum geometry\textcolor{black}{-driven transport} associated with topologically-protected surface states and set the stage for embedding the quantum metric in topological devices.


\begin{figure*}
\centering
\includegraphics[width=1\textwidth]{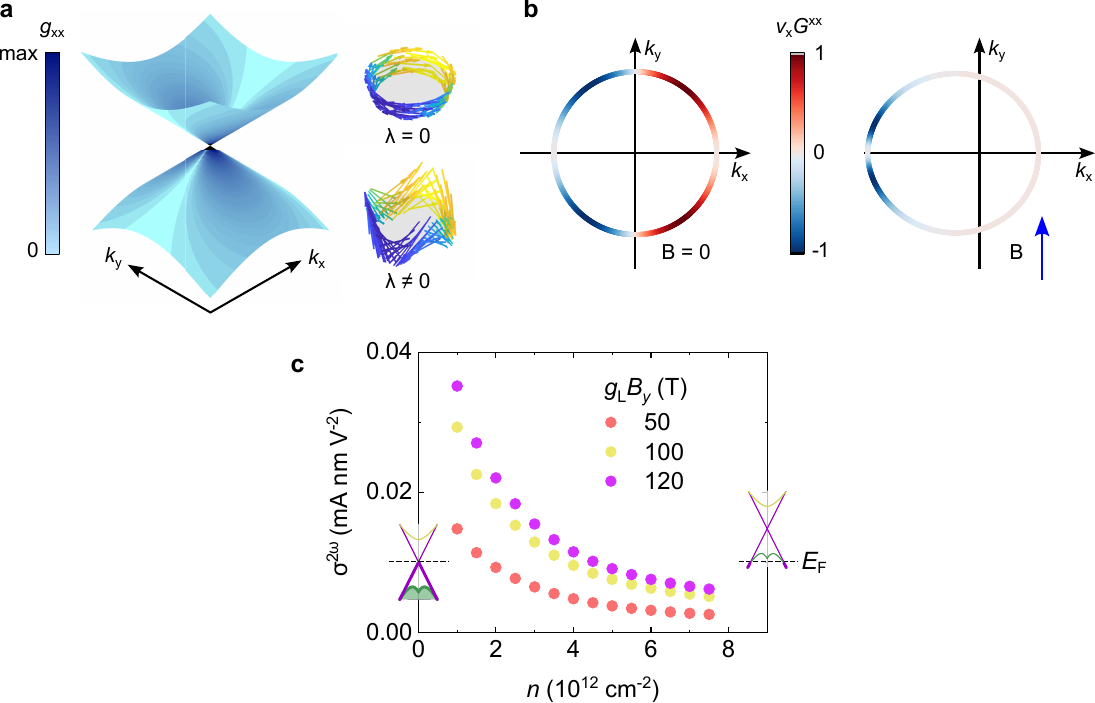}
\caption{\textbf{Quantum geometry of a 3D topological insulator.} \textbf{a}, Left: Surface Dirac cone and quantum metric (diagonal component $g_{xx}$) of a 3D topological insulator with trigonal warping $\lambda = 0$. Right: spin-momentum locking of the Dirac states in the case $\lambda = 0$ and $\lambda \neq 0$. \textbf{b}, Fermi line with $\lambda = 0$, with and without a magnetic field $\mathbf{B} \parallel \hat{y}$. The color codes for the band-energy normalized quantum metric $G^{xx}$ multiplied by the band velocity $v_x = \frac{\partial\epsilon}{\partial k_x}$, scaled to the maximum value. \textcolor{black}{\textbf{c}, Calculated dependence of the quantum metric-driven nonlinear conductivity on the surface carrier density at three effective magnetic fields $g_{\textrm{L}}B_y$, with $g_{\textrm{L}}$ the Landé factor. The warping parameter is $\lambda = 340$ eV\r{A}$^3$. The two sketches show the position of the Dirac cone (in violet) and bulk states (in green) relative to the the Fermi level $E_{\textrm{F}}$ in the low-doping (left) and high-doping (right) regimes.}}
\label{fig1}
\end{figure*}

\section*{The quantum metric of Dirac cones}
The surface states of 3D topological insulators are made of an odd number of Dirac cones with linear energy-momentum dispersion relation captured by the low-energy model $\mathcal{H} = \hbar v_{\textrm{F}}(k_x\sigma_y - k_y\sigma_x) = \textcolor{black}{\mathbf{h(k)}\cdot\sigma}$, where $\mathbf{k}$ and $\mathbf{\sigma}$ are the electron's crystal momentum and spin Pauli matrices, respectively, and $v_{\textrm{F}}$ is the Fermi velocity \cite{Liu2010}. The coupling between $\mathbf{k}$ and $\mathbf{\sigma}$ locks the spin orthogonal to the momentum giving rise to a momentum-space spin texture with a quantum geometric structure (Fig. 1a) \cite{Sala2025,Mercaldo2024}. \textcolor{black}{Similar to Rashba bands at spin-orbit coupled interfaces \cite{Sala2025}, the Dirac cone and its spin-momentum locking feature a nonzero quantum metric tensor $g_{ij} = \frac{1}{4}\partial_{k_i}\hat{\mathbf{h}}\cdot\partial_{k_j}\hat{\mathbf{h}}$. Here, $\hat{\mathbf{h}}$ is the unit vector of $\mathbf{h(k)}$, and $i = x,\, y$. The components of $g$} diverge at the Dirac point. This holds true even in the presence of a warping of the Fermi surface described by the higher order coupling between $\mathbf{k}$ and $\mathbf{\sigma}$ of the form $\frac{\lambda}{2}(k_+^3+k_-^3)\sigma_z$, with $k_{\pm} = k_x \pm ik_y$ and $\lambda$ the coupling strength. This correction is allowed by the trigonal symmetry characterizing the members of the second generation of 3D topological insulators, e.g., Bi$_2$Te$_3$ and Sb$_2$Te$_3$, and leads to a warping of the Fermi contour and spin texture \cite{Fu2009}. 

The quantum metric tensor $g$ of the Dirac cone can be probed by second-order nonlinear transport as long as the time-reversal symmetry is lifted \cite{Kaplan2022,Lahiri2023,Jiang2025}. In time-reversal conditions, nonlinear charge responses associated with the quantum metric are precluded because the dipole of the band-energy normalized quantum metric $G^{aa}$ ($a = x, y, z$) integrates to zero over the Fermi contour (Fig. 1b). However, breaking the time reversal by applying a magnetic field in the surface plane leads to a finite second-order nonlinear conductivity 
$\sigma_{aaa}=\frac{j_{a}^{2\omega}}{(E_{a}^{\omega})^2} \propto \int d^2 {\bf k} ~\partial_{k_{a}} G^{aa} f_0({\bf k})$
in the direction $\hat{a}$ orthogonal to the magnetic field (Methods)  \cite{Mercaldo2024}. Here, $E_{a}^{\omega}$ is the applied electric field, $j_{a}^{2\omega}$ is the ensuing nonlinear current density, and 
$f_0({\bf k})$ is the equilibrium Fermi-Dirac distribution function of the surface band in question.
Microscopically, this happens because the magnetic field shifts the Dirac cone in momentum space and, 
in conjunction with symmetry-allowed particle-hole symmetry breaking terms $\propto k^2$ and the trigonal warping $\propto k^3$,
generates 
a net band-energy normalized quantum metric dipole~\cite{Mercaldo2024}. \textcolor{black}{Thus, the magnetic field does not change the quantum metric $g$, at least at low surface carrier densities, but only the momentum-space integral of the derivatives of $G^{aa}$.}
In comparison, the nonlinear conductivity in the direction parallel to the magnetic field remains zero. Therefore, the concerted action of the quantum metric and magnetic field 
activates a nonlinear and nonreciprocal resistance that can be used as a proxy for the geometrical properties of the Dirac cone. \textcolor{black}{As shown in Fig. 1c, this nonlinear response increases monotonically with the decreasing surface carrier density, which can be modulated, e.g., by chemical doping or gate voltages.} We 
emphasize that, as pointed out in Ref.~\cite{Mercaldo2024}, 
the quantum metric is a general feature of topological surface states described by the Dirac model \cite{Chen2024c,Mera2022}. However, the quantum-metric driven magnetoresistance (QMMR) 
can become finite only
in the presence of quadratic $\sim k^2$ or cubic $\sim k^3$ terms, i.e., the Dirac cone must feature a parabolic correction or a trigonal warping \cite{Mercaldo2024}. These $k$-nonlinear terms, however, are typical of realistic 3D topological insulators \cite{Liu2010}, and the QMMR is thus an effective probe of their quantum metric. \textcolor{black}{We also stress that breaking both the inversion and time-reversal symmetries is necessary to activate second-order nonlinear transport driven by the quantum metric \cite{Kaplan2022,Lahiri2023,Jiang2025}. While in $\mathcal{PT}$-symmetric topological antiferromagnets such as MnBi$_2$Te$_4$ the time reversal is lifted microscopically by the antiferromagnetic order \cite{Gao2023,Wang2023c}, the time reversal symmetry of 3D topological insulators such as Sb$_2$Te$_3$ can be removed by applying magnetic fields \cite{Sala2025,Mercaldo2024}.}

\section*{Quantum metric magnetoresistance}

\begin{figure*}
\centering
\includegraphics[width=1\textwidth]{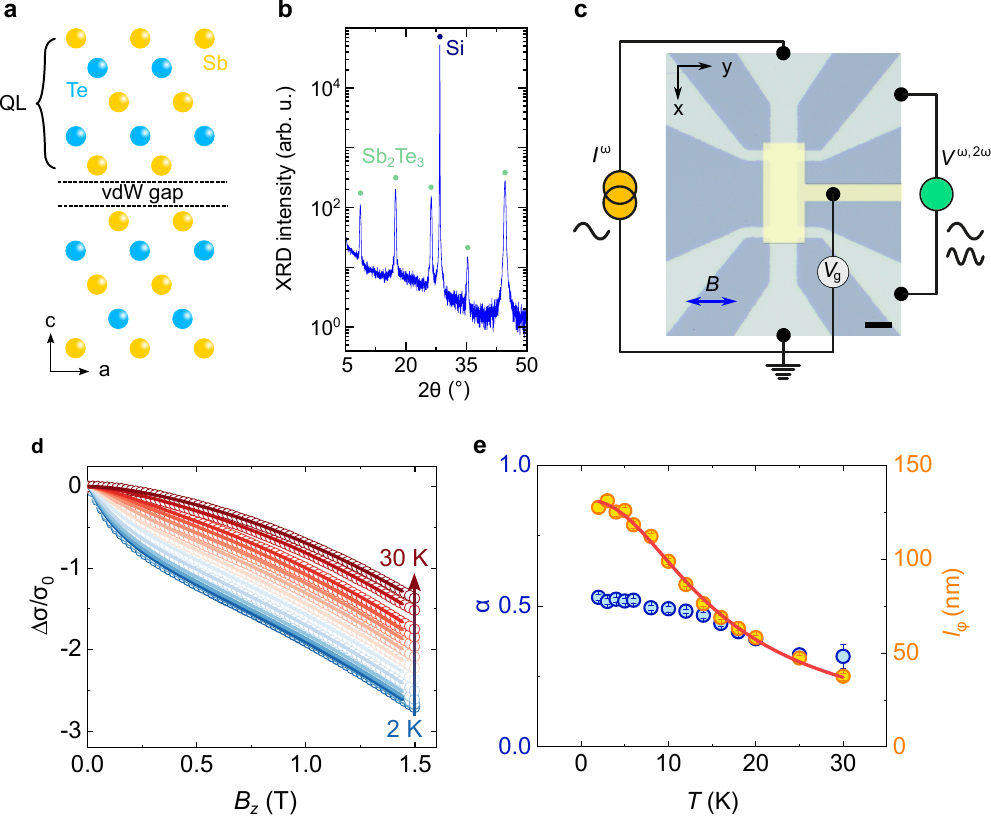}
\caption{\textbf{Sb$_2$Te$_3$ topological insulator.} \textbf{a}, Schematic atomic structure of Sb$_2$Te$_3$ quintuple layers (QL) separated by van der Waals (vdW) gaps. \textbf{b}, X-ray diffraction (XRD) of Sb$_2$Te$_3$(30 nm) grown on Si(111). \textbf{c}, Sketch of the device geometry and nonlinear transport measurement configuration \textcolor{black}{before the fabrication of the top electrode (Methods). The yellow shade represents the top electrode}. The scale bar corresponds to 20 \textmu m. \textbf{d}, Magnetoconductance of Sb$_2$Te$_3$(30 nm) normalized to the quantum of conductance $\sigma_0 = e^2/\pi h$ at increasing temperature. The solid lines are fits of Eq. \ref{eq:WAL} to the data (Methods). \textbf{e}, Weak antilocalization prefactor $\alpha$ and coherence length $l_{\phi}$ extracted from \textbf{d}. The solid line is a fit of Eq. \ref{eq:phaseCoherence} to the data (Methods).}
\label{fig2}
\end{figure*}

To probe the QMMR, we have considered Sb$_2$Te$_3$ thin films  grown by metal-organic chemical vapor deposition (Methods). Sb$_2$Te$_3$ is a chalcogenide-based narrow band-gap semiconducting material formed by quintuple layers connected through van der Waals bonds (Fig. 2a) \cite{Zhang2009}. These films are characterized by a rhombohedral crystalline structure that belongs to the R-3m spatial group and present a nearly epitaxial structure \cite{Rimoldi2020,Shafiei2025}. Figure 2b displays the X-ray diffraction pattern of a 30 nm thick Sb$_2$Te$_3$ film. Intense and sharp peaks associated to the (00\textit{l}) Sb$_2$Te$_3$ planes are observed, in compliance with the high structural order of the material \cite{Rimoldi2020,Anderson1974,Shafiei2025}. The topological insulator character of \textcolor{black}{the films used in this study} has previously been demonstrated by combining angle resolved photoemission spectroscopy (ARPES) and magnetotransport measurements \cite{Locatelli2022,Shafiei2025}, and has been exploited for efficient spin-charge conversion \cite{Longo2021,Longo2022}. \textcolor{black}{ARPES performed on a twin of the Sb$_2$Te$_3$ film considered here identifies linearly dispersing surface bands with the Fermi level located} approximately 0.1 eV below the Dirac point, revealing its p-type character, \textcolor{black}{but does not provide evidence for additional trivial surface states close to the chemical potential}. Furthermore, temperature dependent transport measurements show a metallic behavior (Extended Data Fig. 1) accompanied by weak antilocalization corrections to the magnetoconductance measured at low temperature (Fig. 2d). \textcolor{black}{The dependence of the magnetoconductance on the direction of the magnetic field indicates that the weak antilocalization is contributed by both bulk and surface states \cite{Locatelli2022}.}
The analysis of this effect yields a weak antilocalization prefactor $\alpha \approx 0.5$, which is indicative of one 2D conductive channel \cite{Hikami1980,Steinberg2011,Locatelli2022,Shafiei2025}, and a phase coherence length in the order of 100 nm that is limited by electron-phonon interactions (Fig. 2e, Methods). Overall, ARPES and linear magnetotransport support the presence of \textcolor{black}{surface Dirac cones with} spin-momentum locking, which are the prerequisite for probing effects induced by the quantum metric.

The quantum metric expected for topological surface states can be detected via nonlinear magnetotransport measurements in Hall bar devices \textcolor{black}{traversed by an alternating current $I^{\omega}_{a}$} (Fig. 2c, Methods). Figure 3a,b shows the nonlinear (second-harmonic, 2$\omega$), longitudinal magnetoresistance $R^{2\omega}_{aaa}(B, I^{\omega}_{a}) = \frac{V^{2\omega}_{aaa}}{I^{\omega}_{a}}$ ($a = x,\, y$) \textcolor{black}{measured in two distinct Sb$_2$Te$_3$(30 nm) devices oriented perpendicular and parallel to the magnetic field $\mathbf{B}$ in the absence of any gate voltages} (Methods). When the field is oriented orthogonal to the current direction, $R^{2\omega}$ grows monotonically with the magnetic field, and the increase is linear in the low field region. The resistance changes sign when the magnetic field is inverted, namely, it is antisymmetric in the field. Its amplitude remains approximately constant up to about 30 K, decreases as temperature is increased, and vanishes above 150 K. This nonlinear response is, instead, absent when the field is parallel to the electric current. The dependence of the nonlinear magnetoresistance on the relative orientation is further confirmed by angular scans \textcolor{black}{performed in a third device} (Fig. 3c), which reveal a sinusoidal variation of the nonlinear signal with the azimuthal angle. Additionally, $R^{2\omega}$ scales linearly with the electric current (Fig. 3d and Extended Data Fig. 2), while the linear magnetoresistance remains constant, and does not depend on the measurement frequency (Extended Data Fig. 3). This phenomenology is compatible with the expected QMMR of topological insulators \cite{Mercaldo2024}, which requires breaking time reversal symmetry, but has usually been associated with the bilinear magnetoelectric resistance (BMER). The BMER is not a geometrical effect but originates from the conversion of spin currents into nonlinear charge currents in trigonally warped surface states \cite{He2018a,He2019} or inhomogeneities of the spin-momentum locking in cubic systems \cite{Fu2022}. Our nonlinear magnetoresistance likely includes a contribution from these semiclassical transport phenomena, yet the temperature \textcolor{black}{and mobility} dependencies of both $R^{2\omega}$ and nonlinear conductivity $\sigma^{2\omega} \propto R^{2\omega}$ (Methods) provide compelling evidence for the presence of important quantum geometrical effects below 30 K, as discussed next. 

\begin{figure*}
\centering
\includegraphics[width=1\textwidth]{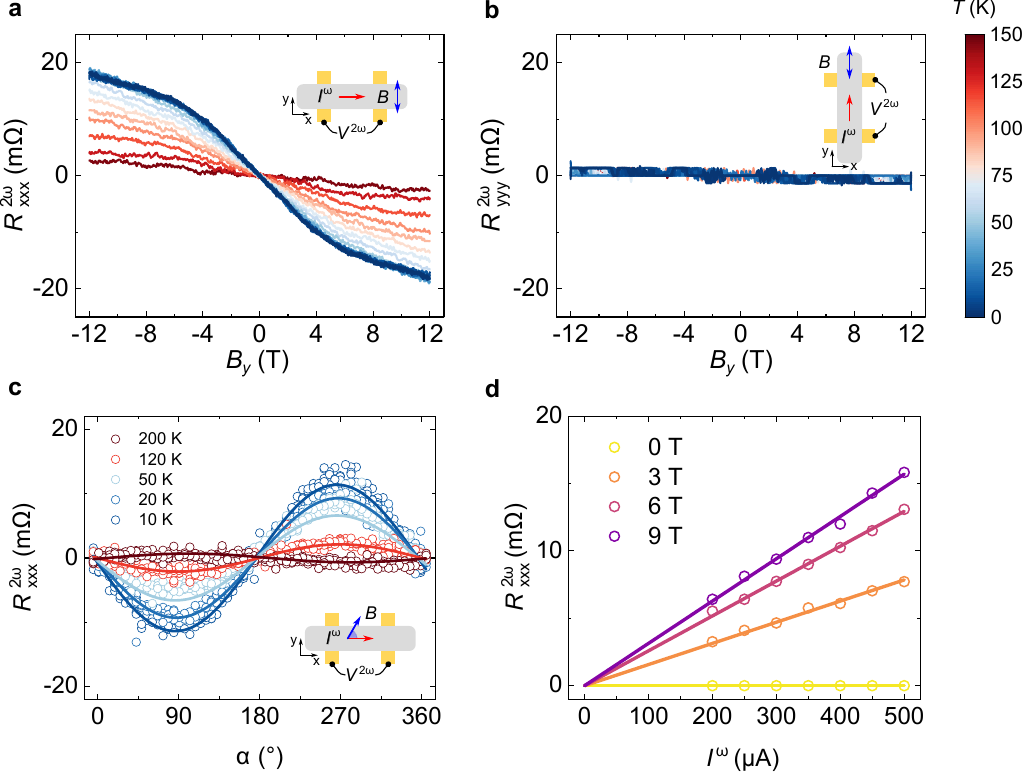}
\caption{\textbf{Nonlinear magnetotransport.} \textbf{a,b}, Nonlinear magnetoresistance as a function of temperature at a current of 500 \textmu A. The orientation of the in-plane magnetic field relative to the current is shown in the sketch of the device. \textbf{c}, Angular dependence of the nonlinear magnetoresistance as a function of temperature at a current of 300 \textmu A \textcolor{black}{and magnetic field of 9 T}. The lines are sinusoidal fits to the data. The azimuthal angle is measured from the direction of the electric current, as shown in the sketch. \textbf{d}, Current dependence of the nonlinear magnetoresistance at 10 K and selected magnetic fields. The lines are linear fits with zero intercept.} 
\label{fig3}
\end{figure*}


In general, $\sigma^{2\omega}$ can be parsed into terms that scale with different powers of the electronic scattering time $\tau$. While the BMER is an extrinsic disorder-mediated effect and, consequently, $\sigma^{2\omega}_{\textrm{BMER}} \sim \tau^2$, the QMMR is intrinsic and, thus, independent of $\tau$, i.e., $\sigma^{2\omega}_{\textrm{QMMR}} \sim \tau^0$ \cite{Kaplan2022,Lahiri2023,Sala2025}. Hall effect measurements indicate that the electronic mobility, hence the scattering time that is proportional to it, increases monotonically upon lowering the temperature (Extended Data Fig. 1). This implies that, if the BMER were preponderant, $R^{2\omega}$ and $\sigma^{2\omega}$ should also become larger with decreasing temperature. In contrast, the nonlinear magnetoresistance and nonlinear conductivity increase as temperature is lowered, but only until a saturation occurring approximately below 30 K, as shown in Fig. 4a,b and Extended Data Fig. 5. This saturation establishes that nonlinear transport at low temperature is independent of $\tau$ and demonstrates, therefore, its quantum geometrical origin. \textcolor{black}{This interpretation is corroborated by a scaling analysis that demonstrates the independence of the nonlinear conductivity from the electronic mobility at temperatures below 30 K (Fig. 4c and Extended Data Fig. 6), as typical of intrinsic effects driven by quantum geometry \cite{Gao2023,Wang2023c}}. Whereas the participation of the BMER in the total signal cannot be excluded, the observation of resistance and conductivity plateaus is a strong indication that the QMMR provides the dominant contribution to the nonlinear transport, at least below 30 K, where the total carrier density is also constant (Extended Data Fig. 1). 

The clear manifestation of the quantum metric only below 30 K can be likely attributed to the loss of spin polarization caused by electronic scattering at higher temperatures. Previous studies of spin transport and spin-orbit torques in topological insulators have indeed proposed that 
electron-electron interactions can be
detrimental to the spin polarization of the surface states \cite{Tang2014,Che2020,Binda2023}. This interpretation was supported by the similar temperature dependencies of the spin polarization and phase coherence length. We observe an analogous trend: the temperature range in which $\sigma^{2\omega}$ remains constant practically coincides with the weak antilocalization regime. This observation suggests that raising the temperature suppresses both the QMMR and weak-antilocalization interference, consistent with their common origin in the spin texture of the topological surface states: as the spin polarization declines, both fade. In addition to this, the reduction of the scattering time, which influences the BMER, and the variation of the carrier density, which affects both the BMER and QMMR, may also contribute to the decrease of $\sigma^{2\omega}$ above 30 K. 
Above this temperature, we cannot therefore evaluate the relative weight of the QMMR and BMER, but the two effects might survive up to 150 K. The decrease of $\sigma^{2\omega}$ with temperature is in line with measurements in other 3D topological insulators \cite{He2018a,He2019}.

\begin{figure*}
\centering
\includegraphics[width=1\textwidth]{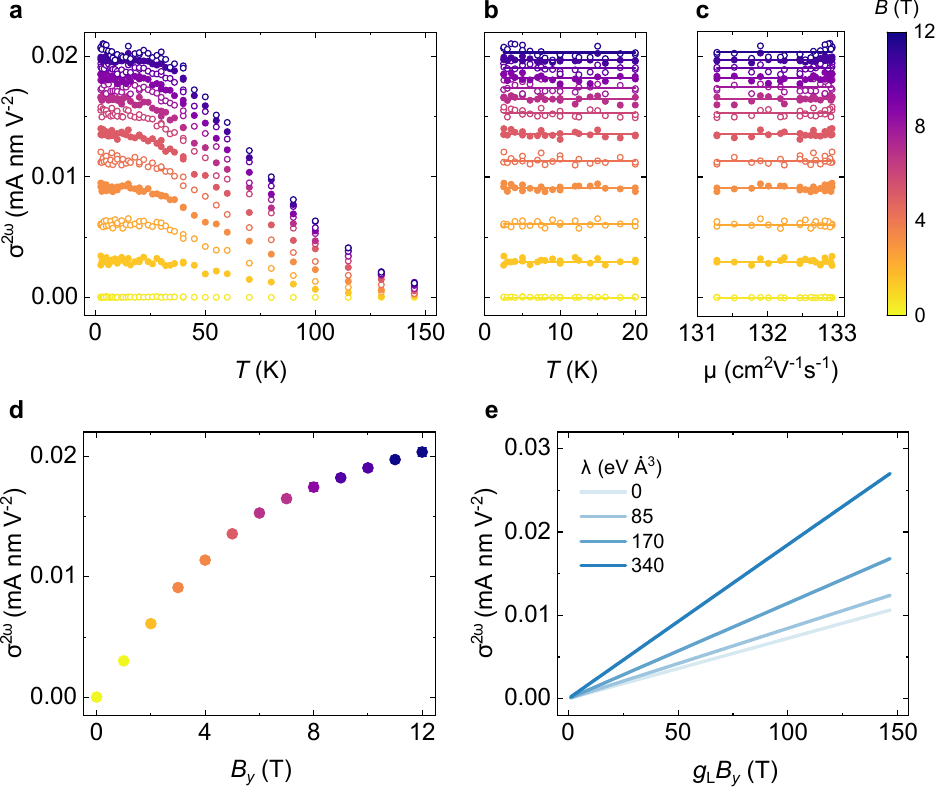}
\caption{\textbf{Quantum-metric nonlinear conductivity.} \textbf{a}, Temperature dependence of the nonlinear conductivity at selected magnetic fields. \textcolor{black}{\textbf{b,c}, Zoom-in on the dependence of the nonlinear conductivity on the temperature and electronic mobility, respectively, in the 2-20 K interval.} The horizontal lines define the average nonlinear conductivity between 2 and 20 K. \textbf{d}, Magnetic field dependence of the averaged nonlinear conductivity between 2 and 20 K. The error bars are as large as or smaller than the size of the data points. \textbf{e}, Calculated dependence of the nonlinear conductivity on the effective magnetic field $g_{\textrm{L}}B_y$ for different values of warping $\lambda$ and a carrier density $n = 2\cdot10^{12}$ cm$^{-2}$. \textcolor{black}{$g_{\textrm{L}}$ is the Landé factor.}} 
\label{fig4}
\end{figure*}

Averaging the nonlinear conductivity in the 2-20 K range allows us to compare our experimental findings with theoretical calculations of the quantum metric contribution to nonlinear transport (Methods) \cite{Mercaldo2024}. As shown in Figs. 4d,e, the calculated quantum metric nonlinear conductivity is of the same order of magnitude as the experimental conductivity. \textcolor{black}{Varying the trigonal warping $\lambda \approx 0-340$ eV\r{A}$^3$ in a large range of values only leads to a factor $\approx 2$ variation of the calculated amplitude of the nonlinear conductivity. This analysis shows that our predictions are robust with respect to large variations of the warping strength. This robustness also holds with respect to uncertainties on the Landé factor $g_{\textrm{L}}$, which is considered to take values in the range $\approx 2-30$ for Sb$_2$Te$_3$ \cite{Liu2010,Korzhovska2020}. If we assume $\lambda = 170$ eV\r{A}$^3$ and take $g_{\textrm{L}} = 2$ ($g_{\textrm{L}} = 30$), our theory model anticipates that $\sigma^{2\omega} = 0.02$ mA nm V$^{-2}$ when $B_y = 75$ T ($B_y = 5$ T), which is larger (smaller) than the 12 T experimental magnetic field at which we do measure $\sigma^{2\omega} = 0.02$ mA nm V$^{-2}$. The comparison with the actual measurements in Fig. 4d suggests, therefore, that the Landé factor is likely in the order of $\approx 10$. The uncertainty on this parameter} as well as finite temperature effects may explain the sublinear dependence of the experimental $\sigma^{2\omega}$ on the magnetic field, which can be theoretically recovered at higher magnetic fields \cite{Mercaldo2024}. However, the good quantitative match between experiment and theory \textcolor{black}{despite parameter uncertainties} corroborates the identification of the low-temperature nonlinear magnetoresistance with the QMMR.

We note that thermoelectric effects cannot explain the saturation of $\sigma^{2\omega}$ at low temperature (Methods). Moreover, as broken inversion symmetry is a necessary condition for nonlinear electronic transport \cite{Kaplan2022,Lahiri2023,Jiang2025}, \textcolor{black}{contributions to $\sigma^{2\omega}$ from the inversion-symmetric bulk are prohibited by symmetry. Thus, the nonlinear resistance that we measure can only stem from the top and bottom surfaces of Sb$_2$Te$_3$. These surfaces host both trivial and topological states. The contribution to $\sigma^{2\omega}$ of spin-split trivial states, however, is very small because of their very large binding energy in Sb$_2$Te$_3$. We estimate from previous ARPES measurements and density functional theory calculations \cite{Pauly2012} that their maximum nonlinear conductivity is of the order of 10$^{-4}$ mA nm V$^{-2}$ (Methods), which is a hundred times smaller than what we measure. This leaves the topological states as the only source of nonlinear transport. In this respect, we note that, although the top-surface and bottom-surface Dirac cones have opposite spin-momentum locking, their contributions to the nonlinear response do not necessarily cancel out. In fact, the different structural and chemical environments of the two interfaces can lead to a relative shift of the top and bottom Dirac cones \cite{hong2012,steinberg2010,yoshimi2015} and, hence, to a net nonlinear response \cite{He2018a,Fu2022}. Gate voltages that selectively target the topological surface states can be instrumental in disentangling these two opposing contributions, as we now demonstrate}.


\textcolor{black}{\section*{Gate voltage control of the quantum metric and QMMR}}

\noindent \textcolor{black}{As reported in Fig. 5a and Extended Data Fig. 4, the voltage $V_{\textrm{g}}$ applied to the top electrode of the Hall bar (Fig. 2c) leads to a few \% variation of the linear resistance $R^{\omega}$ of the device. Given the high bulk carrier density of our samples (Extended Data Fig. 1), only the top-surface Dirac cone is affected by the gate voltage. This interpretation is in line with the observation that increasing $V_{\textrm{g}}$ leads to an increase of the resistance, which corresponds to a shift of the top-surface Dirac point towards the Fermi level (Fig. 5b). Concomitant with the increase of $R^{\omega}$ with $V_{\textrm{g}}$, we also observe a $\approx$ 45\% decrease of the nonlinear resistance $R^{2\omega}$ (Fig. 5a and Extended Data Fig. 4) and, consequently, of the nonlinear conductivity $\sigma^{2\omega} \propto \frac{R^{2\omega}}{(R^{\omega})^3}$. This finding carries two important implications. First, it implies that the measured nonlinear resistance is contributed with opposite sign by both the top and bottom topological surface states, but the bottom Dirac cone provides the largest contribution. This is because, if the top surface was dominating, increasing $V_{\textrm{g}}$, namely, shifting the top-surface Dirac point towards the Fermi level, should lead to an increase of the nonlinear response (Fig. 1c). As we observe the opposite, we conclude that doping the top surface with electrons reduces the energy shift between the top and bottom Dirac cones (Fig. 5b) \cite{chen2010}. Second, these results demonstrate that the gate voltage is an effective knob to tune the net quantum metric of 3D topological insulators and associated response functions. As reported in Extended Data Figs. 5-6, the nonlinear resistance measured at $V_{\textrm{g}} = -40$ V exhibits the same plateaus in temperature and mobility as the nonlinear resistance measured in the ungated topological insulator, thus confirming the intrinsic quantum geometrical origin of the nonlinear response observed below 30 K.}
\newline

\begin{figure*}
\centering
\includegraphics[width=1\textwidth]{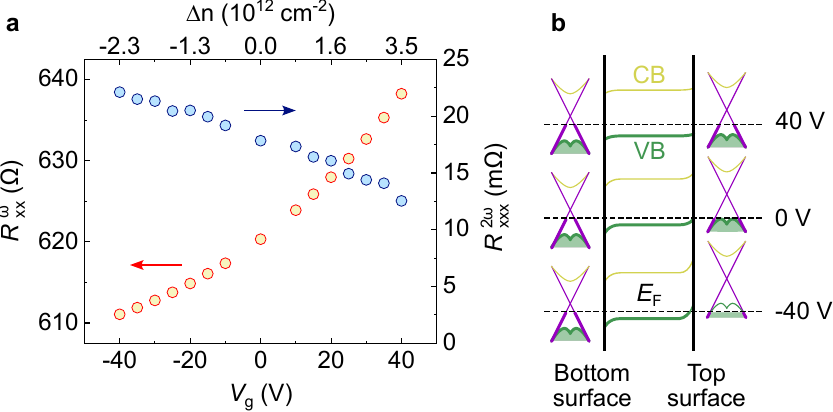}
\caption{\textcolor{black}{\textbf{Gate voltage control of linear and nonlinear transport in Sb$_2$Te$_3$.} \textbf{a}, Gate voltage dependence of the linear magnetoresistance (left) and nonlinear magnetoresistance (right) at a magnetic field $B = 0$ T and $B = 12$ T, respectively, at a temperature $T =$ 2.5 K and with an electric current $I^{\omega} =$ 400 uA. The variation of the 2D carrier density with respect to zero gate voltage is shown on the top $x$ axis, as estimated from the ordinary Hall effect. \textbf{b}, Sketch of the bottom and top surface Dirac cones and surface band bending as a function of the top-electrode voltage. $E_{\textrm{F}}$ is the Fermi level, VB is the valence band, and CB is the conduction band. The surface Dirac cones are in violet, and the bulk states are in green.}} 
\label{fig5}
\end{figure*}

\section*{Discussion}
Our work reveals that the spin-momentum-locked surface states of 3D topological insulators are endowed with a nonzero quantum metric that can be detected via measurements of the nonlinear magnetoresistance \textcolor{black}{and can be controlled with gate voltages}. While this nonlinear response was uniquely attributed to semiclassical effects in previous studies \cite{He2018a,He2019,Fu2022}, our measurements disclose a hitherto unforeseen contribution of geometric origin. As spin-momentum locking is a common feature of topological surface states, a nonzero quantum metric is expected on general grounds in all \textcolor{black}{n-type and p-type} 3D topological insulators. The nonlinear magnetoresistance driven by the quantum metric thus adds to the transport toolbox a new probe of topological surface states and their spin polarization \cite{Bardarson2013}. The quantum metric of 3D topological insulators may also enable new functionalities 
in topological devices
\cite{Mera2022,Breunig2021}, \textcolor{black}{such as nonlinear spin currents \cite{Feng2024,Wang2025b}, photocurrents \cite{Ma2023a}, and second harmonic generation \cite{Bhalla2022}. Because the quantum metric depends on the position of the Fermi level, these effects could be harnessed by tuning the chemical potential by substrate selection \cite{Rimoldi2021,Longo2023}, chemical doping \cite{Ren2012}, and electrical gating, as demonstrated in our study}. 

\bibliography{references}


\section*{Methods}
\noindent\textbf{Sample growth and device fabrication} \\
The Sb$_2$Te$_3$(30 nm) thin films were deposited at room temperature by metal-organic chemical vapor deposition (MOCVD) on 4-inch intrinsic Si(111) wafers (resistivity $>$ 10,000 $\Omega$$\cdot$cm) using an AIXTRON 200/4 system \cite{Rimoldi2020}. This system operates with ultra-high purity nitrogen as the carrier gas and features a cold-wall horizontal deposition chamber fitted with a 4-inch infrared-heated graphite susceptor. The growth of nearly-epitaxial Sb$_2$Te$_3$ layers was optimized over the full 4-inch scale, and their topological character was verified by combining magnetotransport and angular-resolved photoemission spectroscopy \cite{Locatelli2022,Rimoldi2020,Shafiei2025}. X-ray diffraction characterization was performed using a Bruker D8 DISCOVER diffractometer equipped with a four motorized axes stage and an X-ray tube with Cu K-$\alpha$ radiation ($\lambda$ = 1.5406 \r{A}). The characterization was performed in the Bragg-Brentano configuration, which is sensitive to reflections coming from crystalline planes parallel to the sample surface. 

Hall bar devices were fabricated by optical lithography and Ar$^+$ ion milling. The Hall bars have a width $W = 20$ \textmu m and a distance between the longitudinal probes $L = 60$ \textmu m. Ti/Au contact pads were deposited by electron beam evaporation. \textcolor{black}{A top-gate electrode was patterned by e-beam lithography, electron evaporation of SiO$_2$(150)/Ti(10)/Au(60) (thicknesses in nm), and lift-off. The resistance between the Hall bar and the top electrode was in excess of G$\Omega$ at temperatures below 100 K.}
\newline

\noindent\textbf{Magnetotransport measurements} \\
Hall bar devices along orthogonal directions were contacted with Al wire bonds. All transport measurements were performed in an Oxford Teslatron cryostat except for the angular scans, which were performed in a Quantum Design PPMS setup. Nonlinear transport measurements were performed with the magnetic field applied in the sample plane. Measurements of the ordinary Hall effect and weak antilocalization were performed with the magnetic field applied in the direction normal to the sample surface. For nonlinear transport measurements, a \textcolor{black}{commercial current source (Keithley 6221A)} and lock-in amplifiers (Zurich Instruments and Stanford Research) were used to apply an alternating current with an amplitude of a few hundreds \textmu A and a frequency of a few tens Hz and to detect the longitudinal first and second harmonic voltages. First harmonic voltages were measured in phase with the driving current while second harmonic signals were measured out of phase. \textcolor{black}{A DC top-gate voltage was applied by a precision sourcemeter (Keithley 2410). We have verified that the nonlinear response is not an artifact caused by a current-induced modulation of the gate voltage by comparing devices with and without top electrode \cite{Kolling2025}.}

\textcolor{black}{We have performed measurements on three different devices patterned on two samples that were grown together. The field-dependent measurements in Fig. 3a,b and 5a were simultaneously performed in two different devices patterned on the same Sb$_2$Te$_3$(30 nm) sample (sample 1). The angular scans in Fig. 3c were performed on a third device patterned on a distinct Sb$_2$Te$_3$(30 nm) sample (sample 2).}
\newline

\noindent\textbf{Analysis of weak antilocalization effects} \\
The normalized magnetoconductance $\Delta \sigma/\sigma_0 = [\sigma(B) - \sigma(0)]/\sigma_0$ with $\sigma_0 = e^2/\pi h$ as a function of the magnetic field $B$ was fitted to the equation \cite{Hikami1980,Steinberg2011,Locatelli2022}

\begin{equation}
    \frac{\Delta \sigma}{\sigma_0} = -\alpha\left[\Psi\left(\frac{1}{2} + \frac{\hbar}{4el_{\phi}^2B} \right) - \ln\left(\frac{\hbar}{4el_{\phi}^2B}\right)\right] - CB^2,
    \label{eq:WAL}
\end{equation}

\noindent where $\alpha$ is the weak antilocalization prefactor, $l_{\phi}$ is the coherence length, $\Psi$ is the digamma function, and $C$ is a coefficient that captures ordinary magnetoresistance effects. The temperature dependence of $l_{\phi}$ was fitted to the equation \cite{Lin2002}

\begin{equation}
    \frac{1}{l_{\phi}^2} = \frac{1}{l_{\phi_0}^2} + AT^p, 
    \label{eq:phaseCoherence}
\end{equation}

\noindent where $l_{\phi_0}^2$ is the zero-temperature coherence length and $A$ is the coefficient assessing the contribution of electron-electron (e-e) or electron-phonon (e-ph) interactions. Because $p \approx 1$ for e-e scattering and $p \approx 2$ for e-ph scattering \cite{xu2014,shrestha2017}, the value $p \approx 2.4$ obtained from the fit of $l_{\phi}^2$ suggests dominant e-ph interactions.\newline

\noindent\textbf{Analysis of the nonlinear magnetoresistance} \\
The nonlinear resistance was calculated from the measured second harmonic voltage $V^{2\omega}$ as $R^{2\omega}_{aaa} = \frac{V^{2\omega}_{aaa}}{I_{a}^{\omega}}$, with $I_{a}^{\omega}$ the alternating current and $a = x, \,y$. An offset was subtracted from the raw voltage signals, \textcolor{black}{which corresponds to the symmetric part in magnetic field of the nonlinear conductivity}. The nonlinear conductivity $\sigma^{2\omega}_{aaa}$ was defined as $j^{2\omega}_{a} = \sigma^{2\omega}_{aaa}(E_{a}^{\omega})^2 = \sigma_{aa}^{\omega}E_{a}^{2\omega}$, where $j$ and $E$ are the current density and electric field, respectively. The longitudinal nonlinear conductivity was therefore obtained from the relation $\sigma^{2\omega}_{aaa} = \frac{\eta^2 R^{2\omega}_{aaa}}{(R^{\omega}_{aa})^3I_{a}^{\omega}}\frac{L^2}{W}$, where $R^{\omega}_{aa}$ is the linear resistance and $\eta$ quantifies the fraction of current flowing in each surface state. $\eta \approx 2.6 \%$ was calculated by comparing the total carrier density yielded by measurements of the ordinary Hall effect (Extended Data Fig. 1) with the surface carrier density estimated from photoemission experiments performed on a twin Sb$_2$Te$_3$(30 nm) film \cite{Locatelli2022}. \textcolor{black}{$\eta \approx 2.6 \%$ accounts for the contribution of one topological surface state (one Dirac cone) to the nonlinear resistance. Considering one topological surface state instead of two is justified in the regime $V_{\textrm{g}} \leq 0$, where the nonlinear transport is dominated by the bottom Dirac cone. While uncertainties on the value of $\eta$ influence our quantitative estimation of $\sigma^{2\omega}_{aaa}$, they do not affect the temperature and mobility dependencies of $\sigma^{2\omega}_{aaa}$. Therefore, the observed saturation of $\sigma^{2\omega}_{aaa}$ at low temperature and high mobility is independent from the precise value of $\eta$.}
\newline

\noindent\textbf{Role of thermoelectric effects} \\
The ordinary Nernst effect caused by the combination of the out-of-plane thermal gradient and in-plane magnetic field can induce a nonlinear resistance similar to the one observed here. Although a contribution of this effect to the total nonlinear resistance cannot be excluded, it cannot explain the low temperature plateau of the nonlinear conductivity for the following reasons. First, the ordinary Nernst effect is expected to increase linearly with the magnetic field, which is different from the measured field dependence of the nonlinear resistance. Second and more important, the saturation of $\sigma^{2\omega}$ below 30 K is incompatible with the ordinary Nernst effect. The ordinary Nernst current density scales as $j_{\textrm{O}} = \sigma_{\textrm{O}}\Delta T \sim \sigma_{\textrm{O}}\frac{(V^{\omega})^2}{R^{\omega}} \sim \sigma_{\textrm{O}}\frac{(E^{\omega})^2}{R^{\omega}}\sim \sigma_{\textrm{O}}\tau (E^{\omega})^2$, with $\tau$ the electronic scattering time and $\Delta T$ the vertical temperature gradient caused by Joule heating. Since in general the thermoelectric coefficient $\sigma_{\textrm{O}}\sim\tau^q$ with $q\geq 0$, then $j_{\textrm{O}} \sim \tau^{q+1} (E^{\omega})^2$. This implies that the conductivity associated with the ordinary Nernst effect increases as the scattering time becomes longer, i.e., as temperature decreases (Extended Data Fig. 1), which is in contrast with the observed saturation of $\sigma^{2\omega}$.
\newline


\noindent\textbf{Calculation of quantum metric nonlinear conductivity}\\ 
We considered the model Hamiltonian \cite{Liu2010}

\begin{equation}
    \mathcal{H} = \hbar v_{\textrm{F}}(k_x\sigma_y - k_y\sigma_x) + c_2(k_x^2+k_y^2) + \frac{\lambda}{2}(k_+^3+k_-^3)\sigma_z + g_{\textrm{L}}\mu_B(B_x\sigma_x + B_y\sigma_y)
\end{equation}

\noindent which, from left to right, includes the Dirac linear dispersion, a quadratic correction \textcolor{black}{accounting for particle-hole symmetry breaking}, the trigonal warping, and the Zeeman interaction. Here, $\mathbf{k}$ and $\mathbf{\sigma}$ are the electron's crystal momentum and spin Pauli matrices, respectively, $k_{\pm} = k_x \pm ik_y$, and $\mathbf{B}$ is the magnetic field. $v_{\textrm{F}}$ is the Fermi velocity, $c_2$ is the coefficient of the quadratic correction, $\lambda$ is the strength of the trigonal warping, $g$ is the Landé factor, and $\mu_{\textrm{B}}$ is the Bohr magneton. We estimated $\hbar v_{\textrm{F}} = 1.7$ eV\r{A} from the photoemission data in Ref. \cite{Locatelli2022} and set $c_2 = -17$ eV\r{A}$^2$ \cite{Liu2010}. We also varied $\lambda$ in the interval 0-340 eV\r{A}$^3$ but did not specify the value of $g_{\textrm{L}}$ because it can range in a wide interval \cite{Liu2010,Korzhovska2020}. 

We calculated the quantum metric nonlinear conductivity as \cite{Mercaldo2024}

\begin{equation}
    \sigma_{aaa}^{2\omega} = \frac{3e^3}{2\hbar} \sum_n \int \frac{d^2\mathbf{k}}{4\pi^2} \, \partial k_a G^{aa}(\mathbf{k})f(\epsilon_n(\mathbf{k})),
\end{equation}

\noindent where $G^{aa}_n$ is the band-energy normalized quantum metric, $\epsilon_n$ is the energy of the $n^{\textrm{th}}$ band, and $f$ is the Fermi-Dirac distribution function. At zero temperature, this expression can be rewritten as an integral on the Fermi contour

\begin{equation}
    \sigma_{aaa}^{2\omega} \propto \int d\phi\int kdk \, G^{aa} \frac{\partial \epsilon}{\partial k_a}\frac{1}{|\partial \epsilon/\partial k|}\delta[k-k_{\textrm{F}}(\phi)].
\end{equation}

\textcolor{black}{The contribution of the spin-split Rashba surface states to $\sigma_{aaa}^{2\omega}$ was estimated by calculating the quantum metric nonlinear conductivity of Rashba bands using the expression valid in the weak magnetic field regime~\cite{Sala2025}  $$|\sigma_{aaa}^{2 \omega}| \simeq \frac{e^3}{\hbar}   \frac{15}{128\pi} \frac{1}{c_m} \frac{g_L \mu_B B}{\alpha_R } \left[\left(\frac{g_L\mu_B B}{\alpha_R}\right)^2-\frac{\epsilon}{c_m}\right]^{-2},$$
with $c_m=\hbar^2/(2 m^{\star})$, $\alpha_R$ the Rashba parameter, and $\epsilon$ the energy of the Rashba Kramers doublet from the Fermi energy. By fitting the Rashba bands calculated using density functional theory calculations for Sb$_2$Te$_3$ \cite{Pauly2012}, we estimated $\epsilon\simeq 0.83$~eV, $c_{m}=\hbar^2/ 2 m^* \simeq 35$~eV~\r{A}$^2$ and $\alpha_{R}\simeq 0.65$~eV~\r{A}. For $g_LB= 100$~T we then obtain $\sigma_{\text{RS}}\sim 4\cdot 10^{-4}$ mA nm V$^{-2}$.}

\backmatter

\section*{Declarations}
\bmhead{Funding}
G.S. acknowledges support from the Swiss National Science Foundation (grant no. PZ00P2\_223542). E.L. acknowledges the financial support from Projects No. PID2023-152225NB-I00 and Severo Ochoa MATRANS42 (No. CEX2023-001263-S) of the Spanish Ministry of Science and Innovation (Grant No. MICIU/AEI/10.13039/501100011033 and FEDER, EU)"), Projects No. TED2021-129857B-I00 and PDC2023-145824-I00 funded by MCIN/AEI/10.13039/501100011033 and European Union NextGeneration EU/PRTR and by project nº 2021 SGR 00445 by the Generalitat de Catalunya. M.C. acknowledges support from PNRR MUR project PE0000023-NQSTI, and by Italian Ministry of University and Research (MUR) PRIN 2022 under the Grant No. 2022LP5K7 (BEAT). R.M. acknowledges the financial support from the PNRR MUR project PE0000023-NQSTI and the SPIGA project funded by the European Union – NextGenerationEU - PNRR - M4C2, Investment Line 1.1 - PRIN 2022 - ID P2022LXNYN. M.T.M. and C.O. acknowledge partial support by the Italian Ministry of Foreign Affairs and International Cooperation
PGR12351 (ULTRAQMAT) and from PNRR MUR Project
No. PE0000023-NQSTI (TOPQIN). This work was supported by the Swiss State Secretariat for Education, Research and Innovation (SERI) under contract no. MB22.00071, by the Gordon and Betty Moore Foundation (grant no. 332 GBMF10451 to A.D.C.), by the European Research Council (ERC).

\bmhead{Availability of data}
The data that support the findings of this study are available via Zenodo at 10.5281/zenodo.16910820 and from the corresponding authors upon reasonable request.
\bmhead{Acknowledgments}
We thank Jean-Marc Triscone for fruitful discussions and Marco Lopes for technical support. We thank Alberto Morpurgo and Ignacio Gutiérrez for support with the device fabrication.
\bmhead{Author contributions}
E.L. proposed the use of Sb$_2$Te$_3$ topological insulator to probe the quantum metric. R.M. coordinated the production of Sb$_2$Te$_3$. G.S. and E.L. fabricated the devices, performed the transport measurements and data analysis with contribution by S.G.. M.T.M. performed the theory calculations with help from M.C. and C.O. A.D.C. supervised the experimental work, and C.O. supervised the
theoretical work. All authors contributed to writing the manuscript with a first presentation provided by G.S.
\bmhead{Competing interests}
The authors declare no competing interests.

\setcounter{figure}{0}
\renewcommand{\figurename}{Extended Data Fig.}

\clearpage
\begin{figure*}[t!]
\centering
\includegraphics[width=1\textwidth]{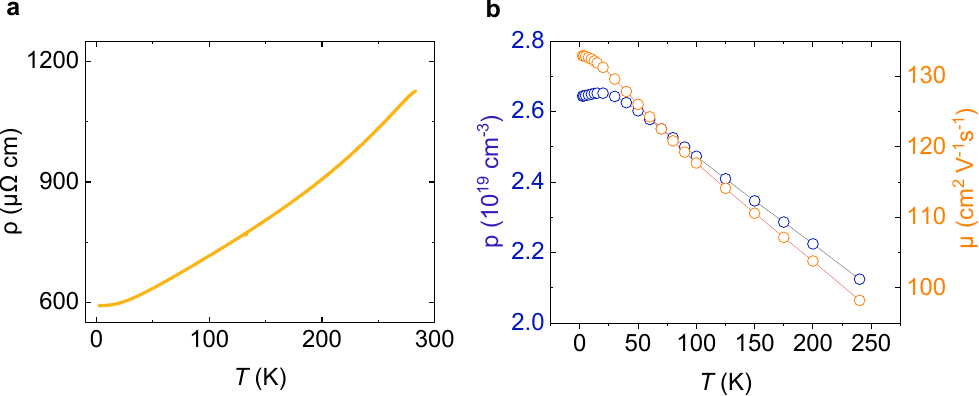}
\caption{\textbf{Linear transport.} \textbf{a}, Temperature dependence of the resistivity of Sb$_2$Te$_3$(30 nm) showing a metallic behavior. \textbf{b}, Temperature dependence of the total (hole) carrier density $p$ and hole mobility $\mu$ estimated from measurements of the ordinary Hall effect.} 
\end{figure*}

\clearpage
\begin{figure*}[t!]
\centering
\includegraphics[width=1\textwidth]{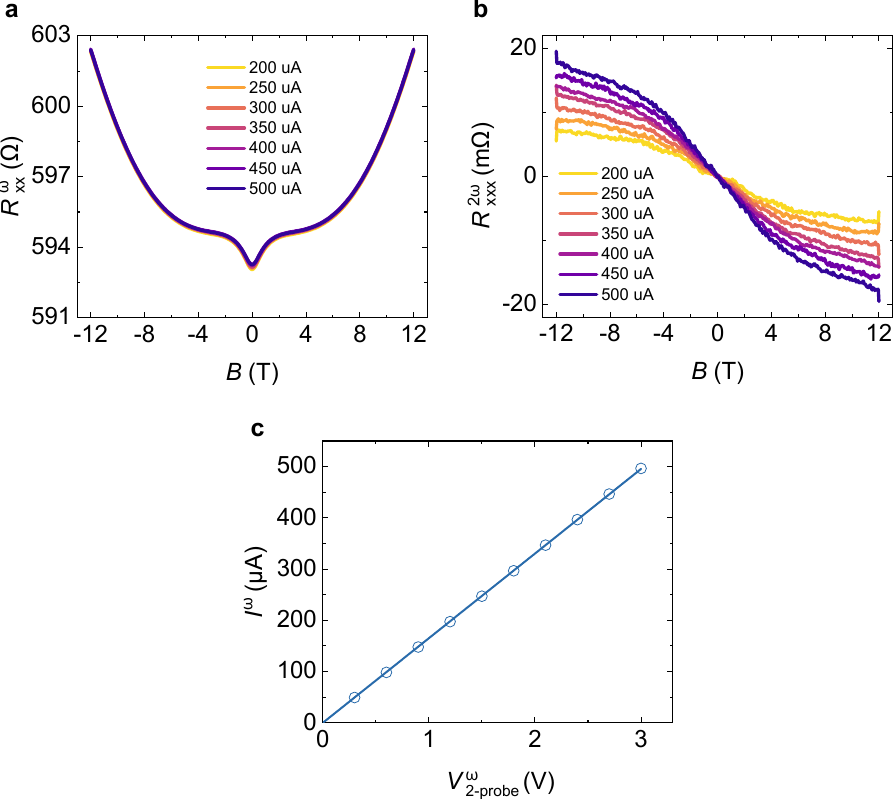}
\caption{\textbf{Current dependence of the linear and nonlinear magnetoresistance.} \textbf{a}, Linear magnetoresistance measured at 10 K at increasing current with the field applied in the sample plane and perpendicular to the current direction. \textbf{b}, Nonlinear magnetoresistance measured simultaneously to the linear magnetoresistance in \textbf{a}. \textbf{c} Two-probe linear I-V curve demonstrating ohmic contacts.} 
\end{figure*}

\clearpage
\begin{figure*}[t!]
\centering
\includegraphics[width=1\textwidth]{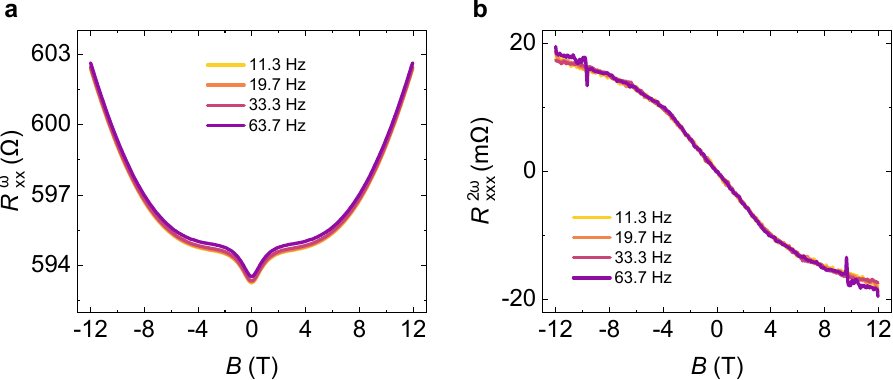}
\caption{\textbf{Frequency dependence} of the linear \textbf{a} and nonlinear \textbf{b} magnetoresistances measured at 10 K with a 500 \textmu A current. The magnetic field was applied in the sample plane and perpendicular to the current direction.}
\end{figure*}

\clearpage
\begin{figure*}[t!]
\centering
\includegraphics[width=1\textwidth]{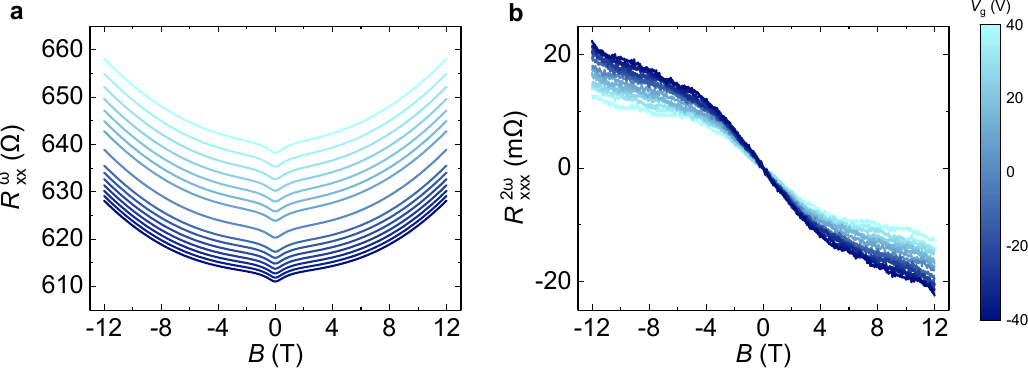}
\caption{\textcolor{black}{\textbf{Dependence of the linear and nonlinear resistance on the top-gate voltage.} \textbf{a,b}, Linear and nonlinear magnetoresistances, respectively, measured with an in-plane magnetic field oriented orthogonal to the current direction as a function of the top-electrode voltage, at a temperature $T =$ 2.5 K and with an electric current $I^{\omega} =$ 400 uA.}} 
\end{figure*}

\clearpage
\begin{figure*}[t!]
\centering
\includegraphics[width=1\textwidth]{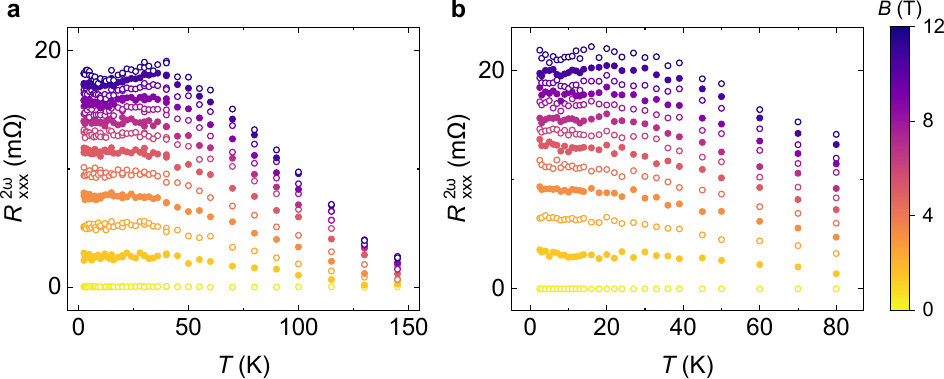}
\caption{\textbf{Temperature dependence} of the nonlinear magnetoresistance at several magnetic fields \textcolor{black} {at gate voltages $V_{\textrm{g}} = 0$ V and $V_{\textrm{g}} = -40$ V in \textbf{a} and \textbf{b}, respectively}.} 
\end{figure*}

\clearpage
\begin{figure*}[t!]
\centering
\includegraphics[width=0.9\textwidth]{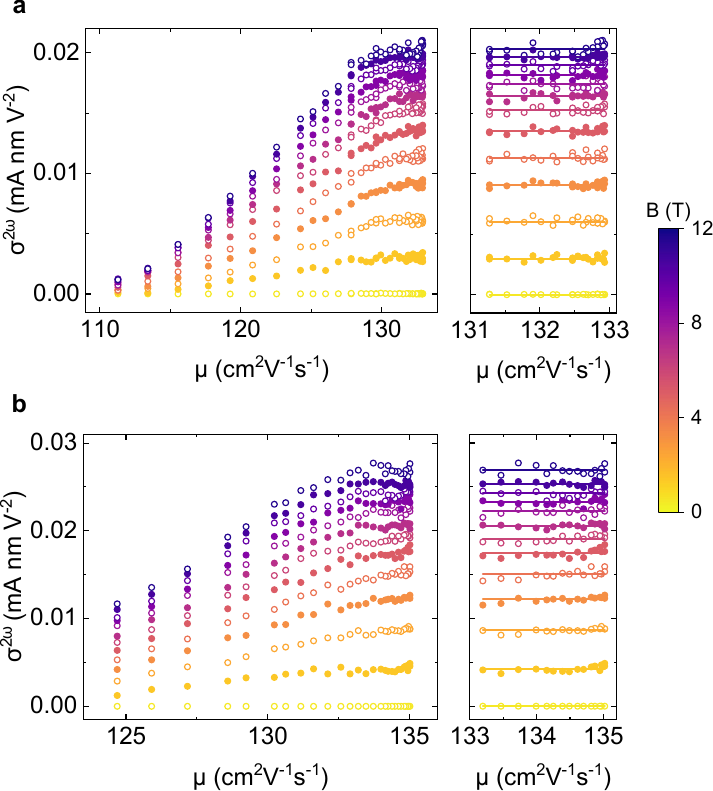}
\caption{\textcolor{black}{\textbf{Quantum-metric nonlinear conductivity as a function of the electronic mobility} at selected magnetic fields and at gate voltages $V_{\textrm{g}} = 0$ V and $V_{\textrm{g}} = -40$ V in \textbf{a} and \textbf{b}, respectively. The graph on the right is a zoom-in on the high-mobility range, which corresponds to temperatures between 2 and 20 K.}} 
\end{figure*}


\end{document}